\documentclass[aps, prd,preprintnumbers,nofootinbib, onecolumn]{revtex4}
\usepackage{bm}
\usepackage{latexsym}
\usepackage{dcolumn}
\usepackage{amsmath,amsfonts,amssymb}
\usepackage{graphicx,epsfig}
\usepackage{color}
\usepackage{amsthm}
\newcommand{\be}{\begin{eqnarray}}
\newcommand{\ee}{\end{eqnarray}}
\newcommand{\bea}{\begin{eqnarray}}
\newcommand{\eea}{\end{eqnarray}}

\begin{document}
%
\title{Effect of the GUP on the Entropy, Speed of Sound, and Bulk to Shear Viscosity Ratio of an ideal QGP}
%

%
%
%
%
\author{Nasser~Demir}\email[email:~]{nasser.demir@ku.edu.kw}
\author{Elias~C.~Vagenas}\email[email:~]{elias.vagenas@ku.edu.kw}
\affiliation{Theoretical Physics Group, Department of Physics, Kuwait University, P.O. Box 5969, Safat 13060, Kuwait}
%
%
%
%
%
%
%
%
%
%
%
\begin{abstract}
\par\noindent
In this work we compute the entropy density, speed of sound, and the resulting impact on the bulk viscosity to 
shear viscosity ratio of an ideal Quark Gluon Plasma when the effects of a generalized uncertainty principle are 
taken into consideration. When the parameter of the generalized uncertainty principle tends to zero, 
i.e., $\alpha \rightarrow 0$, we obtain the value of the speed of sound for the ideal gas of massless particles, 
i.e., $c^{2}_{s}\rightarrow 1/3$, and we recover the expected result that the bulk viscosity $\zeta \rightarrow 0$ 
when $\alpha \rightarrow 0$. In addition, in the high temperature limit, i.e., $T\rightarrow \infty$, the speed of 
sound satisfies the equation  $c^{2}_{s}\rightarrow 1/4$.  The consequence this has on the bulk viscosity 
is that in the high temperature limit, the ratio of the bulk to shear viscosity $\zeta/\eta \rightarrow 5/48$.  Our results suggest that the GUP introduces a scale into the system breaking the a priori conformal invariance of a system of massless noninteracting particles.
\end{abstract}
%
%
\maketitle
%
%
%
%
\section{Introduction}
%
%
%
%
%
\par\noindent
Nowadays it is well known  that General Relativity and Quantum Mechanics are incompatible. From the beginning, it was clear that the Heisenberg Uncertainty Principle (HUP) has to be modified in order to include gravitational effects \cite{GUPearly}. Therefore, the several candidate theories: string theory, loop quantum gravity, 
deformed special relativity, as well as black hole studies provided generalizations of HUP known as 
Generalized Uncertainty Principle (GUP) \cite{amati, kempf, maggiore, fabio}. There exist many ``versions" of the GUP characterized by a GUP parameter $\alpha$. This parameter can be specified either theoretically (see Ref. \cite{amati}), or phenomenologically (see for instance Ref. \cite{elias}).
\par
In the last five years, it was shown that the thermodynamical quantities of an ideal Quark Gluon Plasma (QGP)
are modified when the GUP is taken into account \cite{naggar2013, Elmashad:2012mq, Abou-Salem:2015nka}.  A QGP is a state of matter in which quarks and gluons are deconfined, and it is predicted that this state of deconfinement can be reached at sufficiently high temperatures or densities.  The early universe is expected to be in a deconfined state \cite{Sanches:2014gfa}, and experiments to create a QGP via ``Little Bangs" by colliding heavy nuclei at ultrarelativistic speeds have been undertaken at the Relativistic Heavy Ion Collider (RHIC), at the Brookhaven National Laboratory (BNL) in the USA, and more recently at the Large Hadron Collider (LHC) at CERN in Switzerland, with evidence suggesting a QGP with properties of a near perfect fluid has been created \cite{RHIC_expt}. As such, quantifying thermodynamic and transport properties of this QGP is of relevance.  In particular, the effect of the GUP on the equation of state of an ideal QGP was studied in Ref. \cite{naggar2013}.  Following this line of research, we now proceed to find the GUP corrections to the entropy density, i.e.,  $s$, and to the speed of  sound, i.e., $c_s$, of an ideal QGP.  The computation of the GUP-modified speed of sound is of interest  {\it per se} but it is also useful for studying the effect on the bulk viscosity coefficient, i.e., $\zeta$, since the connection between the speed of sound and the bulk viscosity coefficient is well known both in the context of strongly coupled gauge plasmas \cite{Buchel:2007mf} as well as of bulk media characterized by a small mean free path dominated by radiative quanta \cite{Weinberg:1971mx}.  
\par\noindent
The rest of the paper is organized as follows. In Sec. II we compute the GUP corrections to the entropy density as well as to the speed of sound of an ideal QGP. In Sec. III, using the GUP-modified speed of sound, we derive the GUP corrections to the relation for the bulk and shear viscosity coefficients as given by Weinberg and compare the result to the GUP corrections to the bulk viscosity bound introduced by Buchel.  Finally, in Sec. IV, our results are briefly presented.  In this work, we shall abide by the convention of natural units, setting $\hbar=c=k_B=1.$
%
%
%
%
%
%
%
\section{Entropy and the Speed of Sound}
%
%
%
%
\par\noindent
It is well known that at very high temperatures (energies), the interactions may be very weak due to the asymptotic freedom,  so the QGP can be treated as an ideal gas of quarks and gluons. However, around the critical temperature, namely $T_c$, the QGP may have a non-ideal behavior. For this case, phenomenological models are used to describe this non-ideal behavior of the QGP. One of the most important phenomenological models is the quasiparticle model which treats the QGP as an ideal gas of noninteracting ``massive" fermions and bosons \cite{Gorenstein:1995vm,Peshier:1995ty,Simji:2013kra}. Based on the above, in this section we will derive the GUP corrections to the entropy density and to the speed of sound of an ideal QGP.
\par\noindent
First we define the speed of sound in terms of  the equation of state as follows \cite{Yagi:2005yb}
\begin{equation}
c_s^2 = \frac{\partial P}{\partial \epsilon} = \frac{\partial ln T}{\partial ln s}
\end{equation}
which can be rewritten as
\be
c_s^2 &=& s \frac{\partial }{\partial s} ln T = \frac{s}{T} \frac{\partial T}{\partial s} ~.
\ee
\par\noindent
Therefore, the speed of sound takes the form
\be
c_s^2 = \frac{\frac{s}{T}}{\frac{\partial s}{\partial T}}~.
\label{sos1}
\ee
At this point, it is useful to write the entropy scaling relation for an 
ideal gas of massless particles, i.e.,  Stefan-Boltzmann gas,
\be
s \sim T^3
\label{entropy1}
\ee
which implies that for a Stefan-Boltzmann gas
\be
\frac{\partial s}{\partial T} = 3 \left(\frac{s}{T} \right)~.
\ee
Therefore,  for an ideal gas of massless particles the speed of sound is
\begin{equation}
c_s^2 = \frac{1}{3}~.
\end{equation}
\par\noindent
Next, we introduce the ``version" of the GUP which we will utilize in this work. One of the most common 
deformations of the HUP, i.e.,  
\be
\left[\hat{x}_i , \hat{p}_j \right]= i \hbar\delta_{ij} ~,
\ee
\par\noindent
is given by the expression  \cite{elias}
\be
[x_i, p_j] = i \hbar\hspace{-0.5ex} \left[  \delta_{ij}\hspace{-0.5ex}
- \hspace{-0.5ex} \alpha\hspace{-0.5ex}  \left( p \delta_{ij} +
\frac{p_i p_j}{p} \right)
+ \alpha^2 \hspace{-0.5ex}
\left( p^2 \delta_{ij}  + 3 p_{i} p_{j} \right) \hspace{-0.5ex} \right]~.
\ee
The effect of the aforementioned  GUP on the relativistic dispersion relation is given by \cite{Majhi:2013koa}
\be
E^2(k) = k^2  \left(1- 2 \alpha k \right)  + M^2
\ee
which in the  $M=0$ limit reads
\be
E(k) = k \left(1- 2 \alpha k \right)^{1/2}~.
\ee
\par\noindent
Now, in order to calculate the speed of sound using Eq.(\ref{sos1}), we first have to calculate the entropy density.  The entropy density can be calculated for the system of an ideal QGP by working in the grand canonical ensemble; the entropy density is related to the grand potential, i.e., $\Omega$, via the expression \cite{Yagi:2005yb}
\be
s = - \frac{1}{V} \left(\frac{\partial \Omega}{\partial T} \right)
\label{entropydensity}
\ee
with  the grand potential $\Omega=\Omega(T,V)$  (we have set the chemical potential equal to zero, namely  
$\mu = 0$) to be given as
\be
\Omega = -T  \ln Z
\label{grandpotential}
\ee
\par\noindent
where $Z$ is the grand partition function of the system.  
Following Ref.  \cite{naggar2013},  the system of QGP (whose grand partition function we denote as $z_{QGP}$)
is treated as  a system comprised of fermionic (quarks), bosonic (gluons), and vacuum subsystems 
(whose grand partition functions are denoted by $z_F$, $z_B$, and $z_V$, respectively), thus
\be
\ln z_{QGP} = \ln z_F + \ln z_B + \ln z_{V} 
\label{partition1}
\ee
or, equivalently,
\be
\left.\frac{\Omega}{V}\right|_{QGP}=\left.\frac{\Omega}{V}\right|_{quarks} + \left.\frac{\Omega}{V}\right|_{gluons}+ 
\left.\frac{\Omega}{V}\right|_{vacuum}~.
\label{grandpotentialQGP}
\ee
\par\noindent
The expressions for the grand potentials of the subsystems of quarks, gluons, and vacuum were calculated in Ref.  \cite{naggar2013}
\bea
\left.\frac{\Omega(T, V, \mu=0)}{V}\right|_{quarks} &=& - \frac{7}{8} \frac{\pi^2}{90} g_{q}T^4 -  g_{q} \alpha_2 T^5\\
\left.\frac{\Omega(T, V, \mu=0)}{V}\right|_{gluons}   &=& -  \frac{\pi^2}{90}  g_{g}T^4  -  g_{g} \alpha_1 T^5 \\
\left.\frac{\Omega(T, V, \mu=0)}{V}\right|_{vacuum}   &=& B
\eea
\par\noindent
where  $g_{q}$ and $g_{g}$ are the effective number of degrees of freedom for quarks and gluons, respectively,
$B$ is the bag constant \cite{Chodos:1974je}, $\alpha_1 = \frac{24\zeta(5)}{\pi^{2}}\,\alpha$, 
$\alpha_2 = \frac{45\zeta(5)}{2 \pi^{2}}\,\alpha $, and $\zeta(5)=\sum\limits_{n=1}^{\infty} \frac{1}{n^5}
\approx 1.03693$. Therefore,   Eq.(\ref{grandpotentialQGP}) now reads
\be
\left.\frac{\Omega(T, V, \mu=0)}{V}\right|_{QGP} = 
-\left( \frac{7}{8}g_{q} +g_{g}  \right)\frac{\pi^2}{90}T^{4} - \left( \alpha_{1}g_{g} +\alpha_{2} g_{q} \right) T^{5} +B~.
\ee
\par\noindent
This implies that the GUP-modified entropy density will be
\be
s=\frac{2 \pi^2}{45}\left( \frac{7}{8}g_{q} +g_{g}  \right) T^3 +
  \frac{15 \zeta(5)}{\pi^2}\left(   \frac{15}{2} g_{q} + 8 g_{g}  \right) \alpha T^4  ~.
\label{entropy2} 
\ee
\par\noindent
It is noteworthy that in the limit $\alpha \rightarrow 0$, the entropy density scales $s \sim T^3$, 
which is the Stefan-Boltzmann relation, i.e., Eq.(\ref{entropy1}), in the absence of GUP correction, as one would expect.
It is easily seen that Eq.(\ref{entropy2}) yields
\bea
\left(\frac{\partial s}{\partial T} \right) & = &\frac{6 \pi^2}{45}\left( \frac{7}{8}g_{q} +g_{g}  \right) T^2 +
  \frac{60 \zeta(5)}{\pi^2}\left( \frac{15}{2} g_{q} +8 g_{g}  \right) \alpha T^3\\
\frac{s}{T} &= & \frac{2 \pi^2}{45}\left( \frac{7}{8}g_{q} +g_{g}  \right) T^2 +
  \frac{15\zeta(5)}{\pi^2}\left(  \frac{15}{2}g_{q} + 8 g_{g}  \right) \alpha T^3
\eea
\par\noindent
and after substituting these equations in Eq.(\ref{sos1}), we obtain the GUP modified speed of sound
\be
c_s^2=\frac{\left[\frac{2 \pi^2}{45}\left( \frac{7}{8}g_{q} +g_{g}  \right)\right] T^2 + 
 \left[ \frac{15\zeta(5)}{\pi^2}\left(  \frac{15}{2}g_{q} + 8g_{g}  \right) \right]\alpha T^3}
{3\left[\frac{2 \pi^2}{45}\left( \frac{7}{8}g_{q} +g_{g}  \right) \right]T^2 +
 4\left[ \frac{15\zeta(5)}{\pi^2}\left(  \frac{15}{2}g_{q} + 8g_{g}  \right)\right] \alpha T^3}~.
\label{sos2}
\ee
\par\noindent
At this point a couple of comments are in order. First,  in the limit $\alpha \rightarrow 0$, 
the speed of sound becomes $c_s^2 \rightarrow \frac{1}{3}$, recovering the value of the  speed of sound 
for an ideal gas of massless particles, namely the Stefan-Boltzmann limit.
Second, in the limit  $T \rightarrow \infty$, the speed of sound becomes $c_s^2 \rightarrow \frac{1}{4}$, 
implying that, in the high temperature limit, GUP introduces  some scale that breaks the expected conformal invariance.
\par\noindent
Let us now proceed to find the infinitesimal GUP corrections to the Stefan-Boltzmann relation for the speed of sound.
For this reason, we rewrite Eq.(\ref{sos2}) as follows
\be
c_s^2 = \frac{\frac{1}{3}\left[1 + \left(\frac{675 \zeta(5)}{2\pi^4}\right)\!\left(\frac{\frac{15}{2}g_{q} + 8 g_{g}}{\frac{7}{8}g_{q} +g_{g}}\right)\alpha T\right]}
{\left[1 + \frac{4}{3}\left(\frac{675\zeta(5)}{2\pi^4}\right)\!\left(\frac{\frac{15}{2}g_{q} + 8g_{g}}{\frac{7}{8}g_{q} +g_{g}}\right)\alpha T\right]}
\ee
which we expand in the limit $\alpha T \ll 1$ and obtain
%
%
%
%
\be
c_s^2 \approx \frac{1}{3} -  \frac{75\zeta(5)}{2\pi^4}  
\left(\frac{\frac{15}{2}g_{q} + 8g_{g}}{\frac{7}{8}g_{q} +g_{g}}\right) \alpha T + \mathcal{O}(\alpha^2 T^2)  ~.
\label{sos3}
\ee
Note that the correction to the Stefan-Boltzmann limit of the speed of sound squared is negative, as expected.  Finally, for an ideal QGP comprised of noninteracting gluons and three flavors of massless quarks, the effective number of degrees of freedom for quarks and gluons are given as  $g_q=36$ and $g_g=16$, respectively.  Therefore, the GUP-modified speed of sound, i.e., Eq.(\ref{sos3}),  for an ideal QGP with noninteracting gluons and three flavors of massless quarks reads
\be
c_s^2 \approx \frac{1}{3} - \frac{5970 \zeta(5)}{19\pi^4} \alpha T~.
\label{sos_idealqgp}
\ee
%
%
%
%
%
\section{Impact of Speed of Sound on Bulk Viscosity}
%
%
%
%
\par\noindent
It would be interesting to see what effects this GUP correction has on the ratio of the bulk viscosity 
to shear viscosity, i.e., $\frac{\zeta}{\eta}$  .
In  Ref. \cite{Buchel:2007mf}, a bulk viscosity bound in strongly coupled gauge plasmas was proposed by Buchel
\be
\frac{\zeta}{\eta} \geq 2 \left(\frac{1}{3}-c_s^2 \right)~.
\label{bulkshear1}
\ee
Note that this bound has a different dependence on the speed of sound from the relation
for the shear and bulk viscosity coefficients derived by Weinberg \cite{Weinberg:1971mx}
\be
\frac{\zeta}{\eta}= 15 \left(\frac{1}{3}-c_s^2 \right)^2~.
\label{bulkshear2}
\ee
\par\noindent
However, it should be stressed that Eq.(\ref{bulkshear1}) was proposed for strongly coupled gauge plasmas, 
while Eq.(\ref{bulkshear2}) was derived for a bulk medium characterized by a very small mean free path for radiative quanta. 
\par\noindent 
Now, by substituting the GUP-corrected speed of sound  in the limit $\alpha T \ll 1$ (see  Eq.(\ref{sos3})),  the bulk viscosity 
bound, namely Eq.(\ref{bulkshear1}),  now reads
\be
\frac{\zeta}{\eta} \geq  \frac{75\zeta(5)}{\pi^4}  
\left(\frac{\frac{15}{2}g_{q} + 8g_{g}}{\frac{7}{8}g_{q} +g_{g}}\right) \alpha T ~.
\label{gupbulk1}
\ee
\par\noindent
Eq.(\ref{gupbulk1}) suggests that the GUP introduces a nonzero minimum for the bulk viscosity 
$\zeta$ at arbitrary nonzero temperature $T$.  In the absence of the GUP correction, i.e., when 
$\alpha \rightarrow 0$, the bound then reduces to $\frac{\zeta}{\eta} \geq 0$.  
For a system exhibiting conformal invariance, we expect $\zeta=0$
 and a system of noninteracting massless particles  would exhibit such a conformal invariance.  
However, a nonzero minimum for the bulk viscosity 
in the presence of a GUP correction at arbitrary nonzero temperature implies that the GUP 
introduces a scale breaking the a priori conformal invariance of the system.
\par\noindent
We now proceed to investigate the GUP modification to the Weinberg formula for bulk viscosity for an ideal QGP.  
Indeed Weinberg's formula, namely  Eq.(\ref{bulkshear2}), yields $\zeta=0$ for $c_s^2=\frac{1}{3}$.  
Substituting the GUP-corrected speed of sound in the limit $\alpha T \ll 1$ (see  Eq.(\ref{sos3})) into 
Weinberg's derivation in Eq.(\ref{bulkshear2}) yields
\be
\frac{\zeta}{\eta} \approx  15 \left(\frac{75\zeta(5)}{2\pi^4}\right)^2  
\left(\frac{\frac{15}{2}g_{q} + 8g_{g}}{\frac{7}{8}g_{q} +g_{g}}\right)^{2} \alpha^{2} T^{2}  ~.
\label{gupbulk2}
\ee
Evaluating Eq.(\ref{bulkshear2}) for an ideal QGP of noninteracting gluons and three flavors of 
massless quarks ($g_g=16$ and $g_q=36$) yields
\be
\frac{\zeta}{\eta} \approx \frac{15}{\pi^8} \left(\frac{5970\zeta(5)}{19}\right)^2 \alpha^2 T^2~.
\label{gupbulk_idealqgp}
\ee
At this point a number of comments are in order. First, it is easily seen in Eq.(\ref{gupbulk_idealqgp}) 
that in the limit $\alpha \rightarrow 0$, the bulk viscosity coefficient tends to zero, i.e.,  $\zeta \rightarrow 0$, 
which is also  the case for an ideal gas of massless particles.   
Second, Eq. (\ref{gupbulk2}) suggests that the GUP might introduce 
a scale into the system breaking conformal invariance for a bulk medium characterized by a small 
mean free path dominated by radiative quanta.  Third, as already mentioned, in the high temperature limit, 
i.e., $T \rightarrow \infty$, the speed of sound becomes $c_s^2 \rightarrow \frac{1}{4}$, hence 
modifying the bulk viscosity bound, i.e.,  Eq.(\ref{bulkshear1}) , to 
\be
\frac{\zeta}{\eta} \geq \frac{1}{6}~.
\label{gupbulk3}
\ee
\par\noindent
However, this lower limit of Eq.(\ref{gupbulk3}) does not correspond to the value predicted by 
Weinberg's  formula, since as $c_s^2 \rightarrow \frac{1}{4}$, Eq.(\ref{bulkshear2}) implies 
$\frac{\zeta}{\eta} \rightarrow \frac{5}{48}$,  therefore in the $T \rightarrow \infty$ limit, 
\be
\left( \frac{\zeta}{\eta}\right)_{Weinberg} < \,\, \left( \frac{\zeta}{\eta}\right)_{Buchel,\,min}~.
\ee
\par\noindent
 At this point, it should be stressed that there is no  contradiction here because Buchel's conjecture 
is for strongly coupled gauge plasmas, whereas in the limit $T\rightarrow \infty$, a QGP is expected to be weakly coupled due to the asymptotic freedom property of QCD.  Furthermore, at high temperatures, 
due to increased thermal activity, the mean free path of the medium is expected to be small, and 
hence Weinberg's formula should be more applicable in that regime than Buchel's conjecture.
%
%
%
%
%
%
%
%
\section{Conclusions}
%
%
%
%
%
\par\noindent
In this work we have calculated the GUP corrections to the entropy density $s$, speed of sound $c_s^2$, and the 
resulting impact on the bulk viscosity to shear viscosity ratio, i.e., $\zeta/\eta$, for an ideal QGP.  
We further evaluated specific corrections to the speed of sound and bulk viscosity for an ideal QGP 
of noninteracting gluons and three flavors of massless quarks.  In calculating the GUP corrections to 
all the three aforementioned quantities, we have successfully demonstrated that in the absence of the 
GUP correction, i.e., $\alpha \rightarrow 0$, the results for an ideal gas of noninteracting 
particles are reproduced,  namely $s \sim T^3$, $c_s^2 \rightarrow  1/3$, and $\zeta \rightarrow 0$.   
Additionally, we  found an interesting property when the high temperature limit, i.e., $T \rightarrow \infty$, is 
taken for the GUP-corrected results.  When $T \rightarrow \infty$, $c_s^2 \rightarrow 1/4$, 
and the resulting impact this has on the bulk viscosity to shear viscosity ratio is that the proposed 
bound by Buchel in Ref. \cite{Buchel:2007mf} gets modified to $\zeta/\eta \geq 1/6$. 
We found evidence that the GUP introduces a scale into the system 
breaking its conformal invariance, and this is supported by the finding that even an infinitesimal 
correction to the ideal QGP away from the $T \rightarrow \infty$ limit yields a nonzero value 
for the bulk viscosity (see Eq.(\ref{gupbulk2}) and Eq.(\ref{gupbulk_idealqgp})).  Furthermore, if 
one treats the medium as having a small mean free path dominated by radiative quanta, as 
Weinberg did in the derivation of his formulae in Ref. \cite{Weinberg:1971mx}, the ratio of the 
bulk viscosity to shear viscosity becomes $\zeta/\eta \rightarrow 5/48$ in the $T \rightarrow \infty$ limit, which does not agree with Buchel's proposed bound for $\zeta/\eta$ from Ref. \cite{Buchel:2007mf} for strongly coupled plasmas.  However, the reason for this is easily explained by the fact that QCD is asymptotically free at high temperatures, and hence the QGP is expected to be weakly coupled in the $T \rightarrow \infty$ limit.  It is obvious that more efforts are needed to investigate the GUP corrections to other 
thermodynamic properties of the QGP  and we hope to return to these issues in a future work.
\par\noindent
Finally, one may wonder if there is a way to detect the above-mentioned corrections  in current or future experiments. We believe that there are two ways to detect the GUP corrections for the ideal QGP. First, one may examine the data from gravitational waves (GWs). In particular, one of the sources of the stochastic gravitational wave background (SGWB), which is a random gravitational background for GWs with no specific sharp frequency component, is the cosmological phase transitions  \cite{Binetruy:2012ze}. These phase transitions will produce low-frequency SGWB signals. Among the phase transitions that produce such signals is that between the QGP and the hadronic gas. Therefore, we expect to be able to ``read" the GUP corrections in the  low-frequency SGWB signals when detected. The detection of these signals can be made by the SKA \cite{SKA,Kramer:2004hd} and PTA \cite{IPTA):2013lea} experiments as well as by the eLISA experiment which is planned to ``run'' in a few years \cite{Klein:2015hvg}. More information about this way of detecting the GUP corrections for the ideal QGP can be found in Ref. \cite{Khodadi:2018scn}. It should be pointed out that a subtle point, and thus a point of concern, is the critical temperature used in the calculations in Ref. \cite{Khodadi:2018scn}. In our paper the deconfinement temperature $T_c$ from lattice simulations of QCD matter corresponds to $T_c \sim 170$ MeV \cite{Karsch:2003jg}, but since we are calculating GUP corrections to an ideal QGP, the temperature to approximate the QGP as ideal would need to be higher than $T_c$, perhaps up to $600$ MeV since at lower temperatures the coupling of QCD is too large to treat the QGP as a near ideal gas. 
Second, one may use data from relativistic heavy ion collisions. In particular, there is the issue of how the bulk to shear viscosity  ratio $\zeta/\eta$ could be extracted, albeit indirectly, from relativistic heavy ion collisions.  Extracting transport coefficients from such experiments is quite nontrivial, since the deconfined QGP phase is transient and the collisions evolve through a highly viscous hadronic phase prior to freezeout; the interplay of both the bulk and shear viscous effects and their evolution through the deconfined and hadronic phase have been highlighted in Ref. \cite{Ryu:2015vwa} and Ref. \cite{Song:2009rh}.  Nonetheless, significant efforts have been pursued in this arena \cite{Shen:2015msa, Heinz:2011kt}.  The ratio $\zeta/\eta$ could be extracted by taking the ratio of the bulk viscosity to entropy density ratio $\zeta/s$ and the shear viscosity to entropy density ratio $\eta/s$, both of which would be extracted indirectly.  
Unmasking the viscous effects of the hadronic phase, which is a critical part of this analysis, could be done via separate methods. Such methods could be data-driven hybrid calculations utilizing viscous hydrodynamics for the deconfined phase and microscopic transport models for the hadronic phase \cite{Bernhard:2016tnd, Bernhard:2018hnz}, phenomenological approaches using a particular ansatz for the bulk viscosity \cite{Noronha-Hostler:2015qmd,Noronha-Hostler:2014dqa}, or infinite equilibriated hadronic matter calculations using the Green-Kubo formalism to calculate $\eta/s$ \cite{Rose:2017bjz,Demir:2008tr}. However, one should note that Green-Kubo calculations for $\zeta/s$ in the hadronic phase have not yet been performed.  Another approach to extract an estimate for $\zeta/\eta$  could be using perturbative QCD at high temperatures where the strong coupling is expected to be small; this approach has been taken in Ref. \cite{Csernai:2006zz} for $\eta/s$ and in Ref. \cite{Arnold:2006fz} for $\zeta/s$.  The temperature to use for estimating the values of $\zeta/s$  and $\eta/s$ would be the temperature extracted from an energy density estimate of the initially thermalized fireball in the QGP formation. One should note that it is reasonable to assume that the QGP formed at the LHC is expected to be more weakly coupled than the QGP formed at the RHIC since the total center of mass energy per nucleon is higher and hence the temperature of the produced QGP at the LHC should be higher than that produced at RHIC.
\begin{acknowledgments}
\par\noindent
N.~Demir thanks S.A.~Bass for useful comments regarding the manuscript  and E.C.~Vagenas  thanks S.~Capozziello for useful discussions and enlightening comments.
\end{acknowledgments}
%
%

%
%
%
%
\end{document}